\documentstyle[11pt]{article}
\input{epsf}
\input{psfig}
\begin{document}
\begin{flushright}
GUTPA/99/08/2\\
\end{flushright}
\vskip .1in
\newcommand{\lapprox}{\raisebox{-0.5ex}{$\
\stackrel{\textstyle<}{\textstyle\sim}\ $}}
\newcommand{\gapprox}{\raisebox{-0.5ex}{$\
\stackrel{\textstyle>}{\textstyle\sim}\ $}}
\newcommand{\lsim}{\raisebox{-0.5ex}{$\
\stackrel{\textstyle<}{\textstyle\sim}\ $}}

\begin{center}

{\Large \bf The Quark Lepton Mass Problem and the Anti-Grand Unification
Model}

\vspace{20pt}

{\bf C.D. Froggatt}

\vspace{6pt}

{ \em Department of Physics and Astronomy\\
 Glasgow University, Glasgow G12 8QQ,
Scotland\\}
\end{center}

\section*{ }
\begin{center}
{\large\bf Abstract}
\end{center}

The fermion mass problem and the ideas of mass
protection are briefly reviewed. The Fritzsch ansatz for the
quark mass matrices and a recent variant, based on a
lightest flavour mixing mechanism in which all the CKM mixing angles
disappear in the chiral symmetry limit of vanishing up and down quark
masses, are discussed. The Anti-Grand Unification Model (AGUT) and
the Multiple Point Principle (MPP) used to calculate the values
of the Standard Model gauge coupling constants in the theory are
described. The application of the MPP to the pure Standard Model
predicts the top quark mass to be $173 \pm 5$ GeV and the
Higgs particle mass to be $135 \pm 9$ GeV. Mass protection
by the chiral quantum numbers of the maximal AGUT gauge group
$SMG \times U(1)_f$ provides a successful fit to the
charged fermion mass spectrum, with an appropriate choice of
Higgs fields to break the AGUT gauge group down to the
Standard Model gauge group (SMG) close to the Planck scale.
The puzzle of the neutrino masses and mixing angles presents a
challenge to the AGUT model and approaches to this problem
are briefly discussed.

\vspace{100pt}

Published in the Proceedings of the Corfu Summer Institute on Elementary
Particle Physics, 1998, J. High Energy Phys. Conf. Proc. corfu98/032.

\thispagestyle{empty}
\newpage

\section{Introduction}
\label{intro}

As I discussed in my talk at the previous Corfu workshop \cite{corfu95}
in 1995, the pattern of observed quark and lepton masses,
their mixing and three generation structure
form the major outstanding problem of
particle physics. The hierarchical structure of the charged
fermion masses, ranging over five orders of magnitude from 1/2 MeV for
the electron to 175 GeV for the top quark, and of the quark weak coupling
matrix elements strongly suggests the existence of physics
beyond the Standard Model (SM).
Furthermore the growing experimental support for the
existence of neutrino oscillations and hence for a non-zero
neutrino mass, from SuperKamiokande and other data,
provides direct evidence for non-Standard Model physics.
So the experimental values of the SM fermion masses and
mixing angles presently provide our best clues to the fundamental
physics of flavour.

A fermion mass term
\begin{equation}
{\cal{L}}_{mass} =  -m \overline{\psi}_L \psi_R + h. c.
\end{equation}
couples together a left-handed Weyl field $\psi_L$ and a right-handed
Weyl field $\psi_R$, which then satisfy the Dirac equation
\begin{equation}
 i\gamma^{\mu} \partial_{\mu} \psi_L = m \psi_R
\end{equation}
If the two Weyl fields are not charge conjugates
$\psi_L \neq (\psi_R)^c$
we have a Dirac mass term and the two fields $\psi_L$ and $\psi_R$
together correspond to a Dirac spinor.
However if the two Weyl fields are charge conjugates
$\psi_L = (\psi_R)^c$
we have a Majorana mass term and the corresponding four component
Majorana spinor has only two degrees of freedom.
Particles carrying an exactly conserved charge $Q$, like the electron,
must be distinct from their anti-particles and can only have
Dirac masses with $\psi_L$ and $\psi_R$ having equal charges
$Q_L = Q_R$. However a neutrino could be a massive Majorana
particle.

The left-handed and right-handed top quark, $t_L$ and $t_R$
carry unequal SM $SU(2) \times U(1)$ gauge charges:
\begin{equation}
Q_L \neq Q_R \qquad \mathrm{(Chiral\  charges)}
\end{equation}
Electroweak gauge invariance protects the quarks and leptons
from gaining a fundamental mass term ($\overline{t_L}t_R$ is
not gauge invariant). This {\em mass protection} mechanism is
of course broken by the Higgs effect, which naturally
generates a
mass for the top quark of the same order of magnitude
as the SM Higgs field vacuum expectation value (vev).
Thus the Higgs mechanism explains why the top quark mass
is suppressed, relative to the fundamental
(Planck, GUT...) mass scale of the
physics beyond the SM, down to the scale of electroweak gauge
symmetry breaking. However the further suppression of the other
quark-lepton masses remains a mystery, which it is natural to
attribute to mass protection by another approximately
conserved (gauge) charge (or charges) beyond the SM,
as discussed in section \ref{chiral}.
In this talk I will appeal to the gauge charges of the
Anti-Grand Unification Theory (AGUT) for this mass protection.
The AGUT model and its connection with the Multiple
Point Principle (MPP) is discussed in section \ref{agut}.
The MPP predictions for the top quark and Higgs particle masses
within the pure SM are then discussed in section \ref{higgsmass}.
The Higgs field sector required to break the AGUT gauge group
down to that of the SM is described in section \ref{choosinghiggs}.
The structure of the quark and charged lepton mass matrices
resulting from AGUT mass protection is
presented in section \ref{mass}. I will then consider the
neutrino mass problem in section \ref{neutrino} and
conclude in section \ref{con}.

However let me begin, in the following section \ref{texture},
by considering the structure of the fermion mass matrices and
some of the ans\"{a}tze suggested by phenomenology.

\section{Mass matrix texture}
\label{texture}

The hierarchical structure of the Standard Model fermion mass spectrum
naturally suggests that the fermion mass matrix elements
have a similar hierarchical structure, each typically having
a different order of magnitude. The smaller elements may
then contribute so weakly to the physical masses and
mixing angles that they can effectively be neglected and
replaced by zero---texture zeros. The best known ansatz
incorporating such a texture zero is the two generation
Fritzsch hermitean ansatz \cite{fritzsch}:
\begin{equation}
M_U =\pmatrix{0   	 & B\cr
	      B^{\ast}   & A\cr}
\qquad
M_D =\pmatrix{0   	     & B^\prime\cr
	      B^{\prime\ast}  & A^\prime\cr}
\end{equation}
The assumed hierarchical structure gives the following conditions:
\begin{equation}
|A| \gg |B| , \qquad |A^\prime| \gg |B^\prime|
\end{equation}
among the parameters.
It follows that the two generation
Cabibbo mixing is given by the well-known
Fritzsch formula
\begin{equation}
\left| V_{us} \right| \simeq
\left|\sqrt{\frac{m_d}{m_s}} -
e^{i\phi} \sqrt{\frac{m_u}{m_c}} \right|
\label{fritzsch1}
\end{equation}
where $\phi = \arg{B^\prime} - \arg{B}$. This relationship
fits the experimental value well, provided
that the phase $\phi$ is close to $\frac{\pi}{2}$.
The generalisation of the Fritzsch ansatz to three generations:
\begin{equation}
M_U =\pmatrix{0  		& C   		& 0\cr
		      C^\ast  		& 0   		& B\cr
		      0  		& B^\ast   		& A\cr}
\end{equation}
\begin{equation}
M_D =\pmatrix{0  		& C^\prime  & 0\cr
		      C^{\prime\ast} 	& 0   		& B^\prime\cr
		      0  		& B^{\prime\ast}  & A^\prime\cr}
\end{equation}
with the assumed hierarchy of parameters:
\begin{equation}
|A| \gg |B| \gg |C|, \qquad |A^\prime| \gg |B^\prime| \gg |C^\prime|
\end{equation}
however leads to an additional relationship
\begin{equation}
|V_{cb}| \simeq
\left| \sqrt{\frac{m_{s}}{m_{b}}} -
e^{-i\phi_{2}}\sqrt{\frac{m_{c}}{m_{t}}} \right|
\label{fritzsch2}
\end{equation}
which is excluded by the data for any value of
the phase $\phi_2$.
Consistency with experiment can, for example, be restored by
introducing a non-zero 2-2 mass matrix element \cite{xing}.

There are several ans\"{a}tze, with texture zeros \cite{rrr}, which
give testable relations between the masses and mixing angles
\cite{corfu95}. Here I will discuss a recent suggestion \cite{lfm},
which predicts all the CKM mixing matrix elements in terms of
quark masses. It is a common belief, due
to the success of
eq.~(\ref{fritzsch1}),
that the smallness of the Cabibbo mixing matrix element $V_{us}$
is due to the lightness of the $u$ and $d$ quarks. However not
only the 1-3 generation mixing $V_{ub}$ but also the 2-3 generation
mixing $V_{cb}$ happen to be small compared to $V_{us}$.
This led us to the idea that all
the other mixings, and primarily the 2-3 mixing, could also be
controlled by the up and down quark masses $m_u$ and $m_d$
and vanishes in the chiral symmetry limit $m_u = m_d = 0$.
Therefore we consider an ansatz in which the
diagonal mass matrix elements for the second and third
generations are practically the same in
the gauge (unrotated) and physical bases.

We propose that the three mass matrices for the Dirac
fermions---the up quarks ($U$ = $u$, $c$, $t$), the down
quarks ($D$ = $d$, $s$, $b$) and charged leptons
($E$ = $e$, $\mu$, $\tau$)---are each hermitian with
three texture zeros of the following form:
\begin{equation}
M_i = \pmatrix{  0        &  a_i      &   0  \cr
	      a_i^{\ast} &   A_i     & b_i  \cr
		0        & b_i^{\ast}& B_i \cr} \qquad
i = U, \ D,\ E
\label{LFM1}
\end{equation}
with the hierarchy
$B_i \gg A_i \sim \left| b_i \right| \gg  \left| a_i \right|$
between the elements. Our ansatz requires the diagonal elements
($A_i$, $B_i$), of the mass matrices
$M_i$, to be proportional to the modulus square of the off-diagonal
elements ($a_i$, $b_i$):
\begin{equation}
\frac{A_i}{B_i} = \left| \frac{a_i}{b_i} \right|^2
\qquad
i = U, \ D,\ E
\label{ABab}
\end{equation}
It follows that the Cabibbo mixing is given by the
Fritzsch formula eq.~(\ref{fritzsch1})
which fits the experimental value well, provided
that the CP violating phase $\phi$ is required to be
close to $\frac{\pi}{2}$. Our most interesting prediction
(with the mass ratios calculated at the electroweak
scale \cite{koide}) is:
\begin{eqnarray}
\left|V_{cb} \right| & \simeq &\left| \sqrt{\frac{m_d}{m_b}} -
e^{i\gamma} \sqrt{\frac{m_u}{m_t}} \right| \nonumber \\
& \simeq & \sqrt{\frac{m_d}{m_b}} = 0.038 \pm 0.007
\label{Vcb}
\end{eqnarray}
in good agreement with the current data
$\left| V_{cb} \right| = 0.039 \pm 0.003$ \cite{parodi}.
If we also take the phase $\gamma = \arg{b_D} - \arg{b_U}$
to be $\frac{\pi}{2}$, the uncertainty in our prediction of
eq.~(\ref{Vcb}) is reduced from 0.007 to 0.004.
Another  prediction for the ratio:
\begin{equation}
\left| \frac{V_{ub}}{V_{cb}} \right| = \sqrt{\frac{m_u}{m_c}}
\end{equation}
is quite general for models with nearest-neighbour
mixing.

An alternative scenario, in which the hermitian mass matrix for the up
quarks is changed to be of the form:
\begin{equation}
M_U = \pmatrix{  0        &  0         &   c_U  \cr
	    		 0  	  &   A_U      & 0  \cr
		c_U^{\ast}        &   0        & B_	U \cr}
\label{LFM2}
\end{equation}
leads to mixing angles
given by the simple and compact formulae:
\begin{equation}
\left| V_{us} \right| \simeq \sqrt{\frac{m_d}{m_s}}
\quad
\left| V_{cb} \right| \simeq \sqrt{\frac{m_d}{m_b}}
\quad
\left| V_{ub} \right| \simeq \sqrt{\frac{m_u}{m_t}}
\end{equation}
While the values of $\left| V_{us} \right|$ and
$\left| V_{cb} \right|$ are practically the same
as in our first scenario and in good agreement with experiment, a new
prediction for $\left| V_{ub} \right|$ (not depending on the
value of the CP violating phase) should allow experiment to
differentiate between the two scenarios in the near
future.

\section{Mass matrix texture from chiral flavour charges}
\label{chiral}

As we pointed out in section 1, a natural resolution to the charged fermion
mass problem is to postulate the existence of some
approximately conserved chiral charges beyond the SM.
These charges, which we assume to be the gauge quantum numbers
in the fundamental theory beyond the SM, provide selection
rules forbidding the transitions
between the various left-handed and right-handed quark-lepton
states, except for the top quark. In order to generate mass
terms for the other fermion states,
we have to introduce new Higgs fields, which break the
fundamental gauge symmetry group $G$ down to the SM group.
We also need suitable intermediate fermion states to
mediate the forbidden transitions, which we take to be
vector-like Dirac fermions with a mass of order the
fundamental scale $M_F$ of the theory. In this way
effective SM Yukawa coupling constants are generated, which
are suppressed by the appropriate product of Higgs field
vacuum expectation values measured in units of $M_F$.

Consider, for example, the model
obtained by extending the Standard Model gauge group
$SMG =
SU(3) \times SU(2) \times U(1)$
with a gauged abelian flavour group $U(1)_f$.
This $SMG \times U(1)_f$ gauge group is broken to SMG
by the vev of a scalar field $\phi_S$ where
$\langle\phi_S\rangle <  M_F$ and $\phi_S$
carries  $U(1)_f$ charge $Q_f(\phi_S)$ = 1.
Suppose further that the $U(1)_f$ charges of
the Weinberg Salam Higgs field and the left- and right-handed
bottom quark fields are:
\begin{equation}
Q_f(\phi_{WS})=0 \qquad Q_f(b_L)=0
\qquad Q_f(b_R)=2
\end{equation}
Then it is natural to expect the generation of a
mass for the $b$ quark of order:
\begin{equation}
\left( \frac{\langle\phi_S\rangle }{M_F}
\right)^2\langle\phi_{WS}\rangle
\end{equation}
via a tree level diagram involving
the exchange of two $\langle\phi_S\rangle$ tadpoles,
in addition to the usual
$\langle\phi_{WS}\rangle$ tadpole,
with two appropriately charged vector-like
fermion intermediate states \cite{fn} of mass $M_F$.
We identify
$\epsilon_f=\langle\phi_S\rangle/M_F$
as the $U(1)_f$ flavour symmetry breaking parameter.
In general we expect mass matrix elements of the form:
\begin{equation}
M(i,j) = \gamma_{ij} \epsilon_{f}^{n_{ij}}\langle\phi_{WS}\rangle
\end{equation}
between the $i$th left-handed and $j$th right-handed
fermion components, where
\begin{equation}
\gamma_{ij} = {\cal O} (1),
\quad n_{ij}= \mid Q_f(\psi_{L_{i}}) - Q_f(\psi_{R_{j}})\mid
\label{eq:mij}
\end{equation}

So the {\em effective\/}
SM Yukawa couplings of the quarks and leptons to the
Weinberg-Salam Higgs field
$y_{ij} = \gamma_{ij}\epsilon_{f}^{n_{ij}}$
can consequently be small even though all
{\em fundamental\/} Yukawa couplings of
the ``true'' underlying theory are of $\cal O$(1).
However it appears \cite{bijnens} not possible to explain the
fermion mass spectrum with an anomaly free set of flavour charges
in an $SMG \times U(1)_f$ model with a single Higgs field
$\phi_S$ breaking the $U(1)_f$ gauge symmetry.
In fact it is possible to produce a
realistic quark-lepton spectrum, but at the expense of
introducing three Higgs fields with relatively prime $U(1)_f$
charges and most of the SM fermions carrying exceptionally
large $U(1)_f$ charges. Another possibility is to introduce
SMG-singlet fermions with non-zero values of the $U(1)_f$ charge to
cancel the $U(1)_f^3$ gauge anomaly (as
in $MSSM \times U(1)_f$ models \cite{ibanezross}, which also use
anomaly cancellation via the Green-Schwarz
mechanism \cite{green-schwarz}).
However we shall consider the alternative of
extending the SM gauge group further---in fact to that
of the anti-grand unification model introduced
in the next section.

We shall take the point of view
that, in the fundamental theory beyond the SM,
the Yukawa couplings allowed by gauge invariance
are all of order unity and, similarly,
all the mass terms allowed by gauge invariance are of
order the fundamental mass scale of the theory---say
the Planck scale. Then, apart from the element
responsible for the top quark mass, the quark-lepton
mass matrix elements are only non-zero due to the
presence of other Higgs fields having
vevs smaller (typically by one order of magnitude)
than the fundamental scale. These Higgs fields will,
of course, be responsible for breaking the fundamental
gauge group $G$---whatever it may be---down to the SM group.
In order to generate
a particular effective SM Yukawa coupling matrix element,
it is necessary to break the symmetry group $G$ by a
combination of Higgs fields with the appropriate
quantum number combination  $\Delta \vec{Q}$. When this
``$\Delta \vec{Q}$'' is different for two matrix elements
they will typically deviate by a large factor.
If we want
to explain the observed spectrum of quarks and leptons in this
way, it is clear that we need charges which---possibly in a
complicated way---separate the generations and, at least
for $t-b$ and $c-s$, also quarks in the same generation.
Just using the usual simple $SU(5)$ GUT charges does not
help because both ($\mu_R$ and $e_R$) and
($\mu_L$ and $e_L$) have the same $SU(5)$ quantum numbers.
So we prefer to keep each SM irreducible representation
in a separate irreducible representation of $G$ and
introduce extra gauge quantum numbers distinguishing
the generations, by adding extra Cartesian-product factors to
the SM gauge group.

\section{Anti-Grand unification model}
\label{agut}

In the AGUT model the SM gauge group is extended in much the
same way as Grand Unified $SU(5)$ is often assumed; it is just
that we assume another non-simple gauge group
$G = SMG^3 \times U(1)_f$, where
$SMG \equiv SU(3) \times SU(2) \times U(1)$, becomes active
near the Planck scale $M_{Planck} \simeq 10^{19}$ GeV. So we
have a pure SM desert, without any supersymmetry,
up to an order of magnitude or so below $M_{Planck}$.
The existence of the $SMG^3 \times U(1)_f$ group means
that, near the Planck scale, each of the three quark-lepton
generations has got its own gauge group and associated
gauge particles with the same structure as the SM gauge group.
There is also an extra abelian $U(1)_f$ gauge boson, giving
altogether $3 \times 8 = 24$ gluons, $3 \times 3 = 9$ $W$'s and
$3 \times 1 + 1 =4$ abelian gauge bosons.

The couplings of the $i$'th proto-generation
to the $SMG_i = SU(3)_i
\times SU(2)_i \times U(1)_i$ group
are identical to those to the SM
group. Consequently we have a charge quantization
rule, analogous to the SM charge quantisation rule
(see eq.~(\ref{SMGiChQu}) below), for each
of the three proto-generation weak hypercharge
quantum numbers $y_i$. For the colourless
particles we have the Millikan
charge quantization of all charges
being integer when measured in units
of the elementary charge unit,
but for coloured particles the charges
deviate from being integer by $-1/3$ of the
elementary charge for quarks and by
$+1/3$ for antiquarks.
This rule can be expressed by introducing the concept of
triality $t$, which characterizes the representation of the
centre of the colour $SU(3)$ group, and
is defined so that
$t=0$ for the trivial representation or
for decuplets, octets and so on,
while $t=1$ for triplet ($\underline{3}$) or anti-sextet
etc. and $t=-1$ for anti-triplet
({$ \underline{\overline{3}}$}) or
sextet etc. Then the rule can be  written in the form
\begin{equation}
Q+t/3 = 0 \qquad ( \mbox{mod} \ 1)
\end{equation}
where $Q$ is the electric charge
$ Q=y/2 + t_3/2$ ($t_3$ is the
third component of the weak isospin,
SU(2), and y is the weak hypercharge).
So we may write this SM charge quantization rule as
\begin{equation}
y/2 + d/2 + t/3 =0 \qquad ( \mbox{mod} \ 1)
\label{SMGiChQu}
\end{equation}
where we have introduced the duality $d$, which is defined to
be $0$ when the weak isospin is integer and $d=1$ when it is half
integer.

At first sight, this $SMG^3 \times U(1)_f$ group with
its 37 generators seems to be just one among many
possible SM gauge group extensions. However, it is
actually not such an arbitrary choice, as it
can be uniquely specified by postulating 4 reasonable
requirements on the gauge group $G \supseteq SMG$.
As a zeroth postulate, of course, we require
that the gauge group extension must contain the Standard Model
group as a subgroup $G \supseteq SMG$.
In addition it should obey the
following 4 postulates:

\begin{center}
 The first two are also valid for $SU(5)$ GUT:
\end{center}

\begin{enumerate}
\item $G$ should transform the presently known (left-handed,
say) Weyl particles into each other.
Here we
take the point of view that we do not look for the whole
gauge group $G$, say, but only for that factor group
$G'$ = $G/H$ which transforms the already known quark and lepton
Weyl fields in a nontrivial way. That is to say, we ask for
the group obtained by dividing out the subgroup $H\subset G$
which leaves the quark and lepton fields unchanged. This
factor group $G'$ can then be identified with its representation
of the Standard Model fermions, i.e. as a subgroup of the
$U(45)$ group of all possible unitary transformations of the
45 Weyl fields for the Standard Model. If one took $G$ to be one
of the extensions of SU(5), such as SO(10) or the E-groups
as promising unification groups, the factor group
$G/H$ would be SU(5) only; the extension parts can be said to
only transform particles that are not in the Standard Model
(and thus could be pure fantasy {\it a priori}).

\item No anomalies, neither gauge nor mixed.
We assume that only straightforward anomaly
cancellation takes place and, as in the SM itself,
do not allow for a Green-Schwarz type anomaly
cancellation \cite{green-schwarz}.

\vspace{2 mm}
\begin{center}
But the next two are rather just opposite to the properties
of the $SU(5)$ GUT, thus justifying the name Anti-GUT:
\end{center}

\item The various irreducible representations of Weyl fields
for the SM group remain irreducible under $G$. This is
the most arbitrary of our assumptions about $G$. It
is motivated by the observation that combining SM
irreducible representations into larger unified
representations introduces symmetry relations between
Yukawa coupling constants, whereas the particle spectrum
does not exhibit any exact degeneracies (except
possibly for the case $m_b = m_{\tau}$). In fact
AGUT only gets the naive $SU(5)$ mass predictions as
order of magnitude relations:
$m_b \approx m_{\tau}$, $m_s \approx m_{\mu}$,
$m_d \approx m_e$.
\item $G$ is the maximal group satisfying the other 3
postulates.
\end{enumerate}

With these four postulates a somewhat complicated
calculation shows that,
modulo permutations of the various irreducible representations
in the Standard Model
fermion system, we are led to our gauge group
$SMG^3\times U(1)_f$.
Furthermore it shows that the SM group is embedded
as the diagonal subgroup of $SMG^3$, as required
in our AGUT model. The AGUT group breaks
down an order of magnitude or so below the Planck
scale to the SM group. The anomaly cancellation constraints
are so tight that, apart from various permutations of the
particle names, the $U(1)_f$ charge assignments are
uniquely determined up to an overall normalisation and
sign convention. In fact the $U(1)_f$ group does not couple to
the left-handed particles or any first generation particles,
and the $U(1)_f$ quantum numbers can be chosen as follows:
\begin{equation}
Q_f(\tau_R) = Q_f(b_R) = Q_f(c_R) = 1
\end{equation}
\begin{equation}
Q_f(\mu_R) = Q_f(s_R) = Q_f(t_R) = -1
\end{equation}

The AGUT group breaks
down an order of magnitude or so below the Planck
scale to the diagonal subgroup of
the $SMG^3$ subgroup (the diagonal subgroup is isomorphic to the
usual SM group).
For this breaking we shall use a relatively complicated
system of Higgs fields with names $W$, $T$, $\xi$, and $S$.
In order to fit neutrino masses as well, we need
an even more complicated system.
It should however be said that,
although at the very high energies
just under the Planck energy each generation has its own
gluons, own W's etc., the breaking makes only one
linear combination of a certain colour combination of gluons
``survive'' down to low energies. So below circa 1/10 of the
Planck scale, it is only these linear combinations that are
present and thus the couplings of the gauge particles---at
low energy only corresponding to these combinations---are
the same for all three generations.
You can also say that the phenomenological gluon is
a linear combination with amplitude $1/\sqrt{3}$ for
each of the AGUT-gluons of the same colour combination.
That then also explains why the coupling constant for the
phenomenological gluon couples with a strength that is $\sqrt{3}$
times smaller than for the AGUT-gluons (see eq.~(\ref{alphaagut}) below)
if, as we effectively assume, the three AGUT
$SU(3)$ couplings were equal to each other.

\begin{figure}
\leavevmode
\centerline{\psfig{file=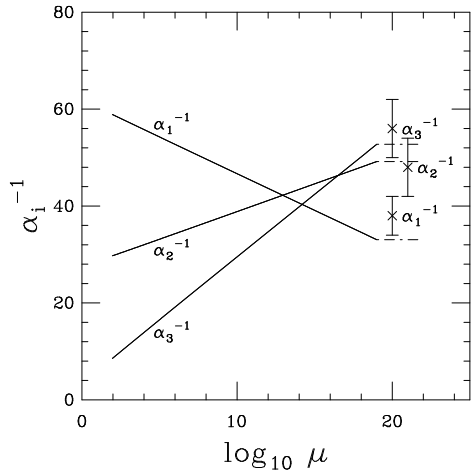,width=6.5cm}
}
\caption{Evolution of the Standard Model
fine structure constants $\alpha_i$ ($\alpha_1$
in the SU(5)
inspired normalisation) from the electroweak scale
to the Planck scale. The anti-GUT model
predictions for the values at the Planck scale,
$\alpha_i^{-1}(M_{Planck})$,
are shown with error bars.}
\label{fig:alphas}
\end{figure}

The SM gauge coupling constants do not,
of course, unify,
because we have not combined
the groups U(1), SU(2) and SU(3)
together into a simple group, but their values have
been successfully calculated using the Multiple Point
Principle \cite{glasgowbrioni}. According to the MPP, the
coupling constants should be fixed such as to ensure the
existence of many vacuum states with the same energy density;
in the Euclideanised version of the theory, there is a
corresponding phase transition. So if several vacua
are degenerate, there is a multiple point. The couplings
at the multiple points have been calculated in
lattice gauge theory for the groups $SU(3)$,
$SU(2)$ and $U(1)$ separately. We imagine that the
lattice has a truly physical significance in providing
a cut-off for our model at the Planck scale. The SM
fine structure constants correspond to those of
the diagonal subgroup of the $SMG^3$ group and,
for the non-abelian groups, this gives:
\begin{equation}
\alpha_i(M_{Planck}) = \frac{\alpha_i^{Multiple \ Point}}{3}
\qquad i=2,\ 3
\label{alphaagut}
\end{equation}
The situation is more complicated for the abelian
groups, because it is possible to have gauge invariant
cross-terms between the different $U(1)$ groups in
the Lagrangian density such as:
\begin{equation}
\frac{1}{4g^2} F_{\mu\nu}^{gen \ 1}(x) F_{gen \ 2}^{\mu \nu}(x)
\end{equation}
So, in first approximation, for the SM $U(1)$ fine
structure constant we get:
\begin{equation}
\alpha_1(M_{Planck}) = \frac{\alpha_1^{Multiple \ Point}}{6}
\end{equation}
The agreement of these AGUT predictions with the data
is shown in figure 1.

\section{The MPP Prediction for the Top Quark and
Higgs masses in the Standard Model}
\label{higgsmass}

\begin{figure}
\leavevmode
\centerline{
\psfig{file=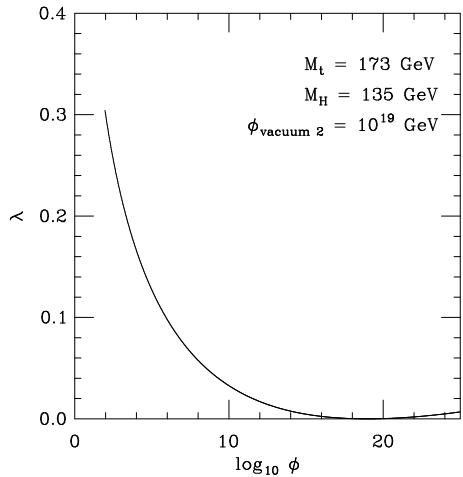,width=6.8cm}}
\caption{Plot of $\lambda$ as a function
of the scale of the
Higgs field $\phi$ for degenerate vacua with the second Higgs
VEV at the Planck scale $\phi_{vac\;2}=10^{19}$ GeV.}
\label{fig:lam19}
\end{figure}
The application of the MPP to the pure Standard Model \cite{fn2}, with
a cut-off close to $M_{Planck}$, implies
that the SM parameters should be adjusted, such that there exists
another vacuum state degenerate in energy density with the
vacuum in which we live. This means that the effective SM
Higgs potential $V_{eff}(|\phi|)$
should, have a second minimum
degenerate with the well-known first
minimum at the electroweak scale
$\langle |\phi_{vac\; 1}| \rangle = 246$ GeV.
Thus we predict that our vacuum is barely stable and we
just lie on the vacuum stability curve in the top quark, Higgs
particle (pole) mass ($M_t$, $M_H$) plane.
Furthermore we expect the second minimum to be within an
order of magnitude or so of the fundamental scale,
i.e. $\langle |\phi_{vac\; 2}| \rangle \simeq M_{Planck}$.
In this way, we essentially select a particular point on
the SM vacuum stability curve and hence the MPP condition
predicts precise values for $M_t$ and $M_H$.

For the purposes of our discussion it is sufficient to consider
the renormalisation group improved tree level effective
potential $V_{eff}(\phi)$.
We are interested in values of the Higgs field
of the order $|\phi_{vac\; 2}| \simeq M_{Planck}$,
which is very large compared to the electroweak scale,
and for which the quartic term
strongly dominates the $\phi^2$ term;
so to a very good approximation
we have:
\begin{equation}
V_{eff}(\phi) \simeq
\frac{1}{8}\lambda (\mu = |\phi | ) |\phi |^4
\end{equation}
The running Higgs self-coupling constant $\lambda (\mu)$
and the top quark running Yukawa coupling constant $g_t(\mu)$
are readily computed by means of the
renormalisation group equations, which
are in practice solved numerically, using the second order
expressions for the beta functions.

\begin{figure}
\leavevmode
\centerline{
\psfig{file=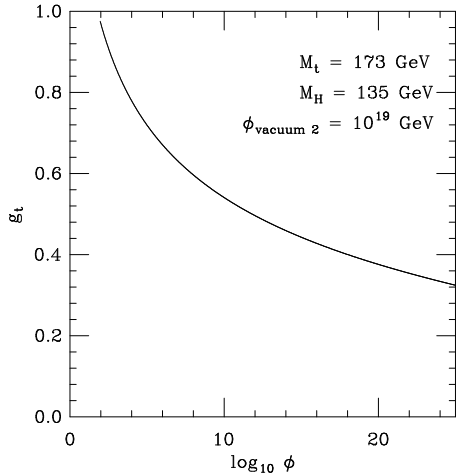,width=6.8cm}
}
\vspace{-0.6cm}
\caption{Plot of $g_t$ as a function
of the scale of the
Higgs field $\phi$ for degenerate vacua with the second Higgs
VEV at the Planck scale $\phi_{vac\;2}=10^{19}$ GeV.}
\label{fig:top19}
\end{figure}

The vacuum degeneracy condition is imposed by requiring:
\begin{equation}
V_{eff}(\phi_{vac\; 1}) = V_{eff}(\phi_{vac\; 2})
\label{eqdeg}
\end{equation}
Now the energy density in vacuum 1 is exceedingly small
compared to $\phi_{vac\; 2}^4 \simeq M_{Planck}^4$. So
we basically get the degeneracy condition, eq.~(\ref{eqdeg}),
to mean that the coefficient $\lambda(\phi_{vac\; 2})$
of $\phi_{vac\; 2}^4$ must be zero with high accuracy.
At the same $\phi$-value the derivative of the effective
potential $V_{eff}(\phi)$ should be zero, because it has
a minimum there.
Thus at the second minimum of the effective potential
the beta function $\beta_{\lambda}$ also vanishes:
\begin{equation}
\beta_{\lambda}(\mu = \phi_{vac\; 2}) =
\lambda(\phi_{vac\; 2}) = 0
\end{equation}
which gives to leading order the relationship:
\begin{equation}
\frac{9}{4}g_2^4 + \frac{3}{2}g_2^2g_1^2 +
\frac{3}{4}g_1^4 - 12g_t^4 = 0
\end{equation}
between the top quark Yukawa coupling and the electroweak
gauge coupling constants $g_1(\mu)$ and $g_2(\mu)$
at the scale $\mu = \phi_{vac\; 2} \simeq M_{Planck}$.
We use the renormalisation group equations to relate the couplings
at the Planck scale to their values
at the electroweak scale.
Figures \ref{fig:lam19} and \ref{fig:top19} show
the running coupling constants $\lambda(\phi)$ and $g_t(\phi)$
as functions of $\log(\phi)$. Their values at the
electroweak scale give our predicted combination of pole
masses \cite{fn2}:
\begin{equation}
M_{t} = 173 \pm 5\ \mbox{GeV} \quad M_{H} = 135 \pm 9\ \mbox{GeV}
\end{equation}

\section{AGUT gauge symmetry breaking by Higgs fields}

\label{choosinghiggs}

There are obviously many different ways to break down the
large group $SMG\times U(1)_f$ to the much smaller SMG. However, we can
greatly simplify the situation by
assuming that, like the quark and lepton fields, the Higgs
fields belong to singlet or fundamental representations of
all non-abelian groups. The non-abelian representations are
then determined from the $U(1)_i$ weak hypercharge quantum
numbers, by imposing the charge quantization rule
eq. (\ref{SMGiChQu}) for each of the $SMG_i$ groups.
So now the four abelian charges, which we express in
the form of a charge vector
\begin{displaymath}
\vec{Q} = \left( \frac{y_1}{2}, \frac{y_2}{2},
\frac{y_3}{2}, Q_f \right)
\end{displaymath}
can be used to specify the complete representation of $G$.
The constraint that we must eventually recover the SM
group as the diagonal subgroup of the $SMG_i$ groups
is equivalent to the constraint that all the Higgs fields
(except for the Weinberg-Salam Higgs field which of course
finally breaks the SMG) should have charges $y_i$ satisfying:
\begin{equation}
\label{diagU1}
y=y_1+y_2+y_3=0
\end{equation}
in order that their SM weak hypercharge $y$ be zero.

We wish to choose the
quantum numbers of the Weinberg-Salam (WS) Higgs
field $\phi_{WS}$ so that it matches the difference in charges between
the left-handed and right-handed physical top
quarks. This will ensure that the top quark
mass in the SM is not suppressed relative
to the WS Higgs field VEV. However we note that
there is a finesse of our fit to the quark-lepton spectrum,
according to which the right-handed component of the experimentally
observed t-quark is actually the one having second
generation $SU(3)$ quantum numbers and is thus really
the proto-right-handed charm quark $c_R$.
In a similar way the
right-handed component of the experimentally observed
charm quark has the third generation $SU(3)$
representation and is really the proto-right-handed top quark $t_R$.
It is only the {\em right}-handed top and charm quarks that
are permuted in this way, while for example the left-handed
components are not. We have to make this identification
of the proto-generation fields $c_R$ and $t_R$; otherwise
we cannot suppress the $b$ quark and $\tau$ lepton masses. This is
because, for the proto-fields, the charge differences
between $t_L$ and $t_R$ are the same as between $b_L$
and $b_R$ and also between $\tau_L$ and $\tau_R$. So
now it is simple to calculate the quantum numbers of
the WS Higgs field $\phi_{WS}$:
\begin{eqnarray}
\vec{Q}_{\phi_{WS}} = \vec{Q}_{c_R} - \vec{Q}_{t_L} &
	=  & \left( 0,\frac{2}{3},0,1 \right) -
	\left( 0,0,\frac{1}{6},0 \right) \nonumber\\
	& = & \left( 0,\frac{2}{3},-\frac{1}{6},1 \right)
\label{ws10}
\end{eqnarray}
This means that the WS Higgs field
will in fact be coloured under both $SU(3)_2$ and
$SU(3)_3$. After breaking the symmetry down to the SM
group, we will be left with the usual WS Higgs field
of the SM and another scalar which will be an octet of
$SU(3)$ and a doublet of $SU(2)$.
This should not present any phenomenological problems,
provided this scalar doesn't cause symmetry breaking
and doesn't have a mass less than about 1 TeV.
In particular an octet of $SU(3)$ cannot lead to baryon
decay.
In our model we take it that what in the Standard Model
are seen as many very small Yukawa-couplings to the
Standard Model Higgs field really represent chain Feynman diagrams,
composed of propagators with Planck scale heavy particles
(fermions) interspaced with
order of unity
Yukawa couplings to
Higgs fields
with the names $W$, $T$, $\xi$, and $S$, which are
postulated to break the AGUT to
the Standard Model Group.
The small effective Yukawa couplings in the
Standard Model are then generated as products of small factors,
given by the ratios of
the vacuum expectation values of
$W$, $T$, and $\xi$
to the masses occurring in the
propagators for the Planck scale fermions in the
chain diagrams \cite{fn}.

The quantum numbers of our invented Higgs
fields $W$, $T$, $\xi$ and $S$
are chosen---and it is remarkable that
we succeeded so well---so
as to make the order of magnitude for the suppressions of the
mass matrix elements of the various mass matrices
fit to the phenomenological
requirements.

After the choice of the quantum numbers for the replacement
of the Weinberg Salam Higgs field in our model, eq.~(\ref{ws10}),
the further quantum numbers needed to be
picked out of the vacuum in
order to give, say, mass to the b-quark is denoted by $\vec{b}$
and analogously for the other particles. For example:


\begin{eqnarray}
\vec{b} & = & \vec{Q}_{b_L} - \vec{Q}_{b_R} - \vec{Q}_{WS} \\
\vec{c} & = & \vec{Q}_{c_L} - \vec{Q}_{t_R} + \vec{Q}_{WS} \\
\vec{\mu} & = & \vec{Q}_{\mu_L} - \vec{Q}_{\mu_R} - \vec{Q}_{WS}
\end{eqnarray}
Here we denoted the quantum numbers of the quarks and leptons
as e.g. $\vec{Q}_{c_L}$ for the left handed components of the
proto-charmed quark.
Note, as we remarked above,
that $\vec{c}$ has been defined using the $t_R$
proto-field, since we have essentially
swapped the right-handed charm and top quarks.
Also the charges of the WS Higgs field
are added rather than subtracted for up-type quarks.

Next we attempted to find some Higgs field quantum numbers which,
if postulated to have ``small'' vevs compared to
the Planck scale masses of the intermediate particles,
would give a reasonable fit to the
order of magnitudes of the mass matrix elements.
We were thereby led to the proposal:

\begin{equation}
\vec{Q}_W = \frac{1}{3}(2\vec{b}+\vec{\mu}) =
		\left( 0,-\frac{1}{2},\frac{1}{2},-\frac{4}{3} \right)
\end{equation}

\begin{equation}
\vec{Q}_T = \vec{b} - \vec{Q}_W =
\left( 0,-\frac{1}{6},\frac{1}{6},-\frac{2}{3} \right)
\end{equation}

\begin{eqnarray}
\vec{Q}_{\xi} = \vec{Q}_{d_L} - \vec{Q}_{s_L} &
	=  & \left( \frac{1}{6},0,0,0 \right) -
	\left( 0,\frac{1}{6},0,0 \right) \nonumber\\
	& = & \left( \frac{1}{6},-\frac{1}{6},0,0 \right)
\end{eqnarray}

{}From the Fritzsch relation \cite{fritzsch}
$V_{us} \simeq \sqrt{\frac{m_d}{m_s}}$
discussed in section \ref{texture},
it is suggested that the two off-diagonal mass
matrix elements connecting the
d-quark and the s-quark be equally big.
We achieve this approximately
in our model by introducing a special Higgs field
$S$, with quantum numbers equal to the
difference between the quantum number
differences for these 2 matrix elements in the
down quark matrix.
Then we postulate that this Higgs field has
a vev of order unity in fundamental units,
so that it does not cause any suppression but
rather ensures that the two matrix elements
get equally suppressed. Henceforth we will
consider the vevs of the new Higgs fields as
measured in Planck scale units and so we have:
\begin{equation}
<S> = 1
\end{equation}
and
\begin{eqnarray}
\vec{Q}_{S} & = & [\vec{Q}_{s_L} - \vec{Q}_{d_R}]
		- [\vec{Q}_{d_L} - \vec{Q}_{s_R}] \nonumber \\
 & = & \left( \frac{1}{6},-\frac{1}{6},0,-1 \right)
\end{eqnarray}
The existence of a non-suppressing
field $S$ means that we cannot
control phenomenologically when this $S$-field is used.
Thus the quantum numbers of the other
Higgs fields $W$, $T$, $\xi$ and $\phi_{WS}$
given above have only been determined modulo those
of the field $S$.

\section{Quark and lepton mass matrices in AGUT}
\label{mass}

We define the mass matrices
by considering the mass terms in the SM to be given by:
\begin{equation}
{\cal L}=\overline{Q}_LM_{U}U_R + \overline{Q}_LM_{D}D_R
+ \overline{L}_LM_{E}E_R+{\rm h.c.}
\end{equation}
The mass matrices can be expressed in terms of the
effective SM Yukawa matrices and the WS Higgs VEV by:
\begin{equation}
M_f = Y_f \frac{<\phi_{WS}>}{\sqrt{2}}
\end{equation}
We can now calculate the suppression factors for
all elements in the Yukawa matrices, by expressing the
charge differences between the left-handed and
right-handed fermions in terms of the
charges of the Higgs fields. They are
given by products of the small numbers
denoting the vevs of the fields $W$, $T$, $\xi$
in fundamental units and the order unity vev of $S$.
In the following matrices we simply write $W$ instead of
$<W>$ etc. for the vevs in Planck units.
With the quantum number
choice given above, the resulting matrix elements
are---but remember that ``random'' complex order
unity factors are supposed to multiply all the matrix
elements---for the uct-quarks:
\begin{equation}
Y_U \simeq \left ( \begin{array}{ccc}
	SWT^2\xi^2
	& WT^2\xi & W^2T\xi \\
	SWT^2\xi^3
	& WT^2 & W^2T \\
	S\xi^3 & 1 & WT
			\end{array} \right ) \label{Y_U}
\end{equation}
the dsb-quarks:
\begin{equation}
Y_D \simeq \left ( \begin{array}{ccc}
	SWT^2\xi^2 & WT^2\xi & T^3\xi \\
	SWT^2\xi & WT^2 & T^3 \\
	SW^2T^4\xi & W^2T^4 & WT
			\end{array} \right ) \label{Y_D}
\end{equation}
and the charged leptons:
\begin{equation}
Y_E \simeq \left ( \begin{array}{ccc}
	SWT^2\xi^2 & WT^2\xi^3
	& S^2WT^4\xi \\
	SWT^2\xi^5 & WT^2 &
	S^2WT^4\xi^2 \\
	S^3WT^5\xi^3 & W^2T^4 & WT
			\end{array} \right ) \label{Y_E}
\end{equation}

We can now set $S = 1$ and fit the nine quark and lepton masses
and three mixing angles, using 3 parameters: $W$, $T$
and $\xi$. That really means we have effectively omitted
the Higgs field $S$ and replaced the maximal AGUT gauge
group $SMG^3 \times U(1)_f$ by the reduced AGUT group
$SMG_{12} \times SMG_3 \times U(1)$, which survives the
spontaneous breakdown due to $S$.
In order to find the best possible fit we
must use some function which measures how
good a fit is. Since we are expecting
an order of magnitude fit, this function
should depend only on the ratios of
the fitted masses to the experimentally
determined masses. The obvious choice
for such a function is:
\begin{equation}
\chi^2=\sum \left[\ln \left(
\frac{m}{m_{\mbox{\small{exp}}}} \right) \right]^2
\end{equation}
where $m$ are the fitted masses and mixing angles and
$m_{\mbox{\small{exp}}}$ are the
corresponding experimental values. The Yukawa
matrices are calculated at the fundamental scale
which we take to be the
Planck scale. We use the first order renormalisation
group equations (RGEs) for
the SM to calculate the matrices at lower scales.

\begin{table}
\caption{Best fit to conventional experimental data.
All masses are running
masses at 1 GeV except the top quark mass
which is the pole mass.}
\begin{displaymath}
\begin{array}{ccc}
\hline
 & {\rm Fitted} & {\rm Experimental} \\ \hline
m_u & 3.6 {\rm \; MeV} & 4 {\rm \; MeV} \\
m_d & 7.0 {\rm \; MeV} & 9 {\rm \; MeV} \\
m_e & 0.87 {\rm \; MeV} & 0.5 {\rm \; MeV} \\
m_c & 1.02 {\rm \; GeV} & 1.4 {\rm \; GeV} \\
m_s & 400 {\rm \; MeV} & 200 {\rm \; MeV} \\
m_{\mu} & 88 {\rm \; MeV} & 105 {\rm \; MeV} \\
M_t & 192 {\rm \; GeV} & 180 {\rm \; GeV} \\
m_b & 8.3 {\rm \; GeV} & 6.3 {\rm \; GeV} \\
m_{\tau} & 1.27 {\rm \; GeV} & 1.78 {\rm \; GeV} \\
V_{us} & 0.18 & 0.22 \\
V_{cb} & 0.018 & 0.041 \\
V_{ub} & 0.0039 & 0.0035 \\ \hline
\end{array}
\end{displaymath}
\label{convbestfit}
\end{table}

We cannot simply use the 3 matrices given by
eqs.~(\ref{Y_U})--(\ref{Y_E}) to calculate
the masses and mixing angles, since
only the order of magnitude of the elements is defined.
Therefore we calculate
statistically, by giving each
element a random complex phase and then
finding the masses and mixing angles.
We repeat this several times and calculate
the geometrical mean
for each mass and mixing
angle. In fact we also vary the magnitude
of each element randomly, by
multiplying by a factor chosen to be
the exponential of a number picked from a
Gaussian distribution with mean value 0 and standard deviation 1.

We then vary the 3 free parameters to
find the best fit given by the $\chi^2$
function. We get the lowest value of $\chi^2$ for the VEVs:
\begin{eqnarray}
\langle W\rangle & = & 0.179   \label{Wvev} \\
\langle T\rangle & = & 0.071   \label{Tvev} \\
\langle \xi\rangle & = & 0.099 \label{xivev}
\end{eqnarray}
The result \cite{smg3m} of the fit is shown
in table~\ref{convbestfit}. This fit has a
value of:
\begin{equation}
\chi^2=1.87
\label{chisquared}
\end{equation}
This is equivalent to fitting 9 degrees of
freedom (9 masses + 3 mixing angles - 3
Higgs vevs) to within a factor of
$\exp(\sqrt{1.87/9}) \simeq 1.58$
of the experimental value. This is
better than might have been
expected from an order of magnitude
fit.

We can also fit to different experimental values
of the 3 light quark
masses by using recent results from lattice QCD, which
seem to be consistently lower than the conventional
phenomenological values.
The best fit in this case \cite{smg3m} is
shown in table~\ref{newbestfit}.
The corresponding values of the Higgs vevs are:
\begin{eqnarray}
\langle W\rangle & = & 0.123	\\
\langle T\rangle & = & 0.079	\\
\langle \xi\rangle & = & 0.077
\end{eqnarray}
and this fit has a larger value of:
\begin{equation}
\chi^2 = 3.81
\end{equation}
But even this is good for an order of magnitude fit.

\begin{table}
\caption{Best fit using alternative light quark masses
extracted from lattice QCD. All masses are running
masses at 1 GeV except the top quark mass
which is the pole mass.}
\begin{displaymath}
\begin{array}{ccc}
\hline
 & {\rm Fitted} & {\rm Experimental} \\ \hline
m_u & 1.9 {\rm \; MeV} & 1.3 {\rm \; MeV} \\
m_d & 3.7 {\rm \; MeV} & 4.2 {\rm \; MeV} \\
m_e & 0.45 {\rm \; MeV} & 0.5 {\rm \; MeV} \\
m_c & 0.53 {\rm \; GeV} & 1.4 {\rm \; GeV} \\
m_s & 327 {\rm \; MeV} & 85 {\rm \; MeV} \\
m_{\mu} & 75 {\rm \; MeV} & 105 {\rm \; MeV} \\
M_t & 192 {\rm \; GeV} & 180 {\rm \; GeV} \\
m_b & 6.4 {\rm \; GeV} & 6.3 {\rm \; GeV} \\
m_{\tau} & 0.98 {\rm \; GeV} & 1.78 {\rm \; GeV} \\
V_{us} & 0.15 & 0.22 \\
V_{cb} & 0.033 & 0.041 \\
V_{ub} & 0.0054 & 0.0035 \\ \hline
\end{array}
\end{displaymath}
\label{newbestfit}
\end{table}

\section{Neutrino mass and mixing}
\label{neutrino}

Physics beyond the SM can generate an effective light neutrino
mass term
\begin{equation}
{\cal L}_{\nu-mass} = \sum_{i, j} \psi_{i\alpha}
\psi_{j\beta} \epsilon^{\alpha \beta} (M_{\nu})_{ij}
\end{equation}
in the Lagrangian, where $\psi_{i,j}$ are the Weyl spinors
of flavour $i$ and $j$, and $\alpha,\beta = 1,2$.
Fermi-Dirac statistics means that the mass matrix $M_{\nu}$
must be symmetric. In models with chiral flavour symmetry we
typically expect the elements of the mass matrices to
have different orders of magnitude. The charged lepton
matrix is then expected to give only a small contribution
to the lepton mixing. As a result of the symmetry of the
neutrino mass matrix and the hierarchy of the mass matrix
elements it is natural to have an almost degenerate pair
of neutrinos, with nearly maximal mixing \cite{fn3}. This
occurs when an off-diagonal element dominates the mass matrix.

A neutrino mass matrix of this texture is generated
in the AGUT model, by tree level diagrams involving
the exchange of two Weinberg Salam Higgs tadpoles and
the appropriate combination of Planck scale Higgs field
tadpoles. The combination which leads to the mass term
$(M_{\nu})_{ij}$ between $\nu_{Li}$ and $\nu_{Lj}$
is determined by the equation
\begin{equation}
\left( \sum \vec{Q}_{\theta} \right)_{ij}
=  \vec{Q}_{\nu Li} + \vec{Q}_{\nu Lj} +
2 \vec{Q}_{\phi_{WS}}
\end{equation}
Here the sum is over the charge vectors for the combination
of Planck scale Higgs fields ($W$, $T$, $\xi$ and $S$)
exchanged. In this way we obtain the neutrino mass matrix
\begin{equation}
\label{eq:Mnuminagut}
M_{\nu} \simeq \frac{\langle {\phi}_{WS} \rangle^2}{M_{Pl}}
\left ( \begin{array}{ccc}
        W^2 {\xi}^4 T^4  & W^2 {\xi} T^4  & W^2 {\xi}^3 T \\
        W^2 \xi T^4  & W T^5 & W^2 T \\
        W^2 {\xi}^3 T  & W^2 T & W^2 T^2 {\xi}^2
\end{array} \right )
\end{equation}
where we have set $<S> = 1$.
The off-diagonal element
$(M_{\nu})_{23} = (M_{\nu})_{32}$
clearly dominates this matrix,
so that we have large mu-tau mixing (between the nearly
degenerate mass eigenstates $\nu_2$ and $\nu_3$).
The mixing matrix $U_{\nu}$ is given
by
\begin{equation}
U_{\nu} \sim \left( \begin{array}{ccc}
1 & \frac{\xi^3}{\sqrt{2}} & \frac{\xi^3}{\sqrt{2}}\\
-\xi^3 & \frac{1}{\sqrt{2}} & \frac{1}{\sqrt{2}}\\
-\xi T^3 & - \frac{1}{\sqrt{2}} & \frac{1}{\sqrt{2}}
\end{array} \right )
\end{equation}
We also have the ratio of neutrino mass squared
differences
\begin{equation}
\frac{\Delta m^2_{23}}{\Delta m^2_{12}} \sim 2 T \xi^2
\sim 1.4 \times 10^{-3}
\end{equation}
giving a hierarchy that is
not suitable for the simultaneous
solution of the solar and atmospheric neutrino problems.

In any case, the mass scale is
much too small to give suitable masses for the atmospheric
neutrino problem. This is because, even if the
$(M_{\nu})_{23}$ element was
unsuppressed by Planck scale Higgs vevs, the see-saw mass
\begin{equation}
\frac{<\phi_{WS}>^2}{M_{Planck}}
\sim 3 \times 10^{-6} \ \mbox{eV}
\label{mseesaw}
\end{equation}
would still be too small.
So, it is necessary to introduce a new mass scale
into the AGUT model in order to obtain
observable neutrino masses and mixings.
This may be done by extending the AGUT Higgs
spectrum to include a weak isotriplet Higgs field
$\Delta$ with SM weak hypercharge
$\frac{y}{2} = -1$. However there is some unnaturalness
in obtaining a value for $<\Delta^0>$ from the scalar
potential some orders of magnitude greater than the
see-saw mass of eq.~(\ref{mseesaw})

Furthermore we need extra structure for the lepton mass matrices
and must relax the assumption that all the
independent matrix elements are of different orders of
magnitude. For example $M_{\nu}$ may have two
order of magnitude degenerate elements $A \sim B$
with a texture of the form:
\begin{equation}
M_{\nu} =
\left(
\begin{array}{ccc}\times & A & B\\
A & \times & \times \\
B & \times & \times \end{array}\right )
\label{Mnu2}
\end{equation}
where $\times$ indicates texture zeros.
The mass eigenvalues are given by:
\begin{equation}
m_{\nu i} = \pm \sqrt{A^2 + B^2}, \, 0, \hspace{1cm} (i = 1, 2, 3)
\end{equation}
although these will be slightly altered when the effects
of the small elements represented by texture zeros
are included. With these eigenvalues we clearly have a hierarchy in
$\Delta m^2$'s with the more degenerate pair being heavier:
\begin{equation}
\Delta m^2_{12} \ll \Delta m^2_{13} \sim \Delta m^2_{23}.
\label{eq:hierarchy}
\end{equation}
So we take $\Delta m^2_{12} = \Delta m^2_{solar}$,
$\Delta m^2_{23} = A^2 + B^2  \sim 10^{-3} \ \mbox{eV}^2$, where
$\Delta m^2_{solar}$ will depend on the type of solution we adopt
for the solar neutrinos.

The corresponding neutrino mixing matrix (assuming that
the charged lepton mass matrix $M_E$ is quasi-diagonal) is:
\begin{eqnarray}
U_{\nu} & \sim & \left( \begin{array}{ccc}
1 & 0 & 0\\
0 & \cos \theta & - \sin \theta \\
0 & \sin \theta & \cos \theta \\
\end{array} \right) \label{eq:typeVmix}
\left( \begin{array}{ccc}
\frac{1}{\sqrt{2}} & -\frac{1}{\sqrt{2}} & 0\\
\frac{1}{\sqrt{2}} & \frac{1}{\sqrt{2}} & 0\\
0 & 0 & 1\\
\end{array} \right) \nonumber \\
& = & \left( \begin{array}{ccc}
\frac{1}{\sqrt{2}} & -\frac{1}{\sqrt{2}} & 0\\
\frac{1}{\sqrt{2}} \cos \theta & \frac{1}{\sqrt{2}} \cos \theta &
	-\sin \theta\\
\frac{1}{\sqrt{2}} \sin \theta & \frac{1}{\sqrt{2}} \sin \theta &
	\cos \theta\\
\end{array} \right)
\label{eq:bi-maxU}
\end{eqnarray}
where
\begin{equation}
\tan \theta = \frac{B}{A}.
\end{equation}
{}From the first row we can see that $\nu_e$ is maximally
mixed between $\nu_1$ and $\nu_2$, so that its mixing does not
contribute to the atmospheric neutrino anomaly, and there
will be no effect observable at Chooz. The atmospheric neutrino
anomaly will be entirely due to large $\nu_{\mu} - \nu_{\tau}$
mixing and, in order that the mixing be large enough, we need
$\sin^2 2\theta \ge 0.8$ ($90 \% C.L$) which requires
\begin{equation}
\label{eq:ABratio}
0.56 \le \frac{B}{A} \le 1.8
\end{equation}
so that, although $A$ and $B$ must be order of magnitude degenerate,
it is not necessary to do any fine tuning. The solar
neutrino problem is explained by vacuum oscillations, although
whether it is an `energy-independent' or a `just-so' solution
will depend on the small elements which we have neglected.
It cannot be explained by an MSW type solution since the mixing
between $\nu_e$ and $\nu_{\mu}$ is too large for this type
of solution, and will remain too large even after the texture
zeroes are removed. The particular case of $B = A$ for this texture
corresponds to the popular bi-maximal mixing solution \cite{bi-maximal}
to the solar and atmospheric neutrino problems.
This type of structure cannot explain the LSND result
and does not give a significant
contribution to hot dark matter, since the sum of the neutrino masses
is given by
\begin{eqnarray}
\sum m_{\nu} & \sim & 2 \sqrt{A^2+B^2} \sim 2
\sqrt{\Delta m^2_{atm}} \nonumber\\
& < & 0.2 \ \mbox{eV}
\end{eqnarray}

We have not been able to extend the Higgs sector of the
AGUT model in such a way as to obtain a neutrino mass matrix
$M_{\nu}$ with the above texture of eq.~(\ref{Mnu2}).  However we
have constructed \cite{smgu2} an anomaly free
Abelian extension of the
Standard Model, which naturally yields a mass matrix $M_{\nu}$
of this type. This $SMG \times U(1)^2$ model was inspired by
the AGUT model and has exactly the same charged fermion
spectrum as in the AGUT fit of Table \ref{convbestfit}.
In order to rescue the AGUT neutrino mass and mixing predictions,
it seems necessary to introduce yet another Higgs field and obtain
the large mixing required for the atmospheric neutrino problem from
the charged lepton mass matrix $M_E$. The solution to the
solar neutrino problem can then be obtained from $M_{\nu}$
or from the mixing due to small elements in $M_E$. This, of course,
has to be achieved without signicantly disturbing the quality of
the AGUT fit to the charged fermion spectrum.

\section{Conclusions}
\label{con}

We emphasized the hierarchical structure of the quark-lepton mass
spectrum and how it points to a mass protection mechanism,
controlled by approximately conserved chiral (gauge) charges
beyond the Standard Model. The structure of ans\"{a}tze for
the fermion mass matrices, suggested by the hierarchy of masses
and mixing angles, was briefly discussed.
A recent ansatz based on a lightest flavour mixing mechanism
was discussed, which gives simple and compact formulae
for all the CKM mixing angles in terms of the quark masses.

The anti-grand unification theory (AGUT), and how the associated
multiple point principle (MPP) is used to predict the values
of the three Standard Model gauge coupling constants, was described.
Applied to the case of the pure Standard Model, the MPP
leads to our predictions
for the top quark and Higgs pole masses:
$M_t = 173 \pm 5$ GeV and $M_H = 135 \pm 9$ GeV.

The AGUT group $SMG^3 \times U(1)_f$
is characterised by being the largest anomaly-free gauge group
acting on just the 45 SM Weyl fermions, without any
unification of the SM irreducible representations.
This group
assigns a unique set of anomaly free chiral gauge
charges to the quarks and leptons. With an appropriate
choice of Higgs field quantum numbers, the AGUT chiral
charges naturally give a realistic charged fermion
mass hierarchy. An order of magnitude fit
in terms of 3 Higgs vevs is
given in Table \ref{convbestfit}, which
reproduces all the masses and mixing angles within a factor
of two. The most characteristic feature of the fit is
that, apart from the $t$ and $c$ quarks, the masses
of the particles in the same generation are predicted to
be degenerate (but only in order of magnitude) at the
Planck scale. The worst feature is the deviation, by a factor of
about 2, between the fitted and experimental values for
$m_s$ and $V_{cb}$.

On the other hand, the puzzle of the neutrino masses
and mixing angles presents a challenge to the model.
It is necessary to introduce a new mass scale into
the AGUT model, using say a weak isotriplet
Higgs field $\Delta$, in order to generate a neutrino mass
appropriate to atmospheric neutrino oscillations.
Using a reduced model, based on the gauge group $SMG \times U(1)^2$,
it is possible to obtain a reasonably natural solution to the
solar and atmospheric neutrino problems and, at the same time,
reproduce the successful AGUT fit to the charged fermion spectrum.
However it is not possible to embed this Abelian extension of
the SM into the AGUT, since one cannot choose a consistent set of
non-Abelian representations for the Higgs fields. It appears
that we shall have
to relax the assumption that the charged lepton mass matrix is
quasi-diagonal, in order to rescue the AGUT model.

\section{Acknowledgements}
\label{ack}
I should like to thank the organisers George Koutsoumbas, Nick Tracas
and George Zoupanos for their hospitality in Corfu.


\begin{thebibliography}{99}
\bibitem{corfu95}
C. D. Froggatt, To be published in the {\it Proceedings of the
5th Hellenic School amd Workshops on Elementary Particle
Physics} (Corfu, 1995);
hep-ph/9603432.
%
\bibitem{fritzsch}
H. Fritzsch, {\it  Phys. Lett.} {\bf B70} (1977) 436;
{\bf B73} (1978) 317.
%
\bibitem{xing}
H. Fritzsch and Z. Z. Xing, {\it  Phys. Lett.} {\bf B338} (1995) 114.
%
\bibitem{rrr}
P. Ramond, R. G. Roberts and G. G. Ross,
{\it Nucl. Phys.} {\bf B406} (1993) 19.
%
\bibitem{lfm}
J. L. Chkareuli and C. D. Froggatt,
{\it  Phys. Lett.} {\bf B450} (1999) 158.
%
\bibitem{koide}
H. Fusaoka and Y. Koide, {\it Phys. Rev.}  {\bf D57} (1998) 3986.
%
\bibitem{parodi}
Particle Data Group, {\it Eur. Phys. J.} {\bf C3} (1998) 1.
%
\bibitem{fn}
C. D. Froggatt and H. B. Nielsen,
{\it Nucl. Phys.} {\bf B147} (1979) 277.
%
\bibitem{bijnens}
J. Bijnens and C. Wetterich,
{\it Nucl. Phys.} {\bf B283} (1987) 237.
%
\bibitem{ibanezross}
L. E. Ibanez and G. G. Ross,
{\it  Phys. Lett.} {\bf B332} (1994) 100.
%
\bibitem{green-schwarz} M. B. Green  and J. Schwarz
{\it Phys. Lett.} B{\bf 149} (1984) 117.
%
\bibitem{glasgowbrioni}D.L. Bennett, C.D. Froggatt and H.B. Nielsen in
{\em Proceedings of the 27th
International Conference on High Energy Physics}
(Glasgow, 1994), eds.~P. Bussey and I. Knowles,
(IOP Publishing Ltd, 1995) p.~557;
{\em Perspectives in Particle Physics '94},
eds. D. Klabu\u{c}ar, I. Picek and D. Tadi\'{c},
(World Scientific, 1995) p.~255, hep-ph/9504294.
%
\bibitem{fn2}
C. D. Froggatt and H. B. Nielsen,
{\it  Phys. Lett.} {\bf B368} (1996) 96.
%
\bibitem{smg3m}
C. D. Froggatt, H. B. Nielsen and D. J. Smith,
{\it  Phys. Lett.} {\bf B385} (1996) 150; \\
C. D. Froggatt, M. Gibson, H. B. Nielsen and D. J. Smith,
{\it Int. J. Mod. Phys.} {\bf A13} (1998) 5037.
%
\bibitem{fn3}
C. D. Froggatt and H. B. Nielsen,
{\it Nucl. Phys.} {\bf B164} (1979) 114.
%
\bibitem{bi-maximal}
M.Jezabek, {\em These Proceedings}.
%
\bibitem{smgu2}
C. D. Froggatt, M. Gibson and H. B. Nielsen,
{\it  Phys. Lett.} {\bf B446} (1999) 256.

\end{thebibliography}
\end{document}